\documentclass[useAMS,usenatbib]{mn2e}
\usepackage[dvips]{graphicx}
\usepackage{url}
\usepackage{amssymb}
\usepackage{times}
\usepackage{subfigure}
\title[Maser lines in IRAS~$15452-5459$]{Probing the fast outflow in IRAS~$15452-5459$ with ATCA observations of OH, H$_2$O, and SiO masers }
\author[L. Cerrigone, K. M. Menten, and H. Wiesemeyer]{L. Cerrigone$^{1}$\thanks{E-mail:
cerrigone@cab.inta-csic.es (LC)}, K. M. Menten$^2$\thanks{E-mail: kmenten@mpifr.de (KMM)} and H. Wiesemeyer$^2$\thanks{E-mail: hwiese@mpifr.de (HW)}\\
$^{1}$Centro de Astrobiolog\'{\i}a, INTA-CSIC, Torrej\'on de Ardoz 28850, Spain\\
$^{2}$Max-Planck-Institut f\"ur Radioastronomie, Bonn 53121, Germany}
\begin{document}
\newcommand{\kms}{km~s$^{-1}$}
\date{}

\pagerange{\pageref{firstpage}--\pageref{lastpage}} \pubyear{2013}

\maketitle

\label{firstpage}

\begin{abstract}
Maser lines of OH, H$_2$O, and SiO are commonly observed in O-rich AGB stars, but their presence after the end of the Asymptotic Giant Branch (AGB) phase is linked to non-spherical mass-loss processes.  IRAS~$15452-5459$ is a post-AGB star with an hourglass nebula whose maser lines are quite peculiar. We observed all of the three maser species with the Australia Telescope Compact Array with angular resolutions of 6$''$, 0.6$''$, 0.3$''$, and 1.7$''$ at 18 cm, 13 mm, 7 mm, and 3 mm, respectively. While double peaks are routinely seen in OH and water masers  and interpreted as due to expanding envelopes, only very few sources display SiO lines with a similar spectral profile. Our observations confirm the detection of the double peak of SiO at 86 GHz; the same spectral shape is seen in the lower-J maser at 43 GHz. A double peak is also detected in the water line, which covers the same velocity range as the SiO masers. Thermally excited lines of SiO are detected at 7 and 3 mm and span the same velocity range as the maser lines of this species. Although observations at higher angular resolution are desirable to further investigate the spatial distributions of the maser spots, the current data allow us to conclude that the SiO masers are distributed in an hourglass shape and are likely due to the sputtering of dust grains caused by shock propagation. The complex OH profile would instead be due to emission from the fast outflow and an orthogonal structure.
\end{abstract}


\section{Introduction}
The evolutionary phase between the end of the Asymptotic Giant Branch (AGB) and the Planetary Nebula (PN) phases is of particular interest in the lives of intermediate-mass stars. 
O-rich stars at the tip of the AGB typically show maser emission from SiO, H$_2$O, and OH molecules that trace different regions of the circumstellar environment. 

In AGB stars, SiO masers arise from  high-density ($>$10$^{9}~$cm$^{-3}$) and high-temperature ($>$1000~K) gas very close to the stellar atmosphere, having as an outer limit the dust condensation region. 
H$_2$O masers arise in regions  with temperatures of 300--1000 K and densities of 10$^7$--10$^9$~cm$^{-3}$ \citep[see Fig. 12, for an illustrative example]{reid1997}. These values correspond approximately to the region of dust formation and wind acceleration. Finally, OH masers arise in the cooler, less dense environment of the outer envelope. 
Assuming spherical symmetry for the mass loss process, when the central star stops ascending the AGB and starts contracting at constant luminosity, the conditions for SiO masers rapidly disappear (a few $\times$ 10 yr after the tip). The subsequent dilution and cooling of the circumstellar envelope (CSE) due to its expansion will make the other masers disappear in a few  $\times$ 100  (H$_2$O) or 1000 (OH) yr \citep{lewis}.


{The detection of OH masers in AGB stars is linked to the optical depth of their surrounding envelope, 
hence to their mass-loss rate. A  sequence can be drawn, going from stars with small mass-loss rates ($\dot{\mathrm M} \lesssim 10^{-7}$~M$_\odot$~yr$^{-1}$), which rarely display any OH maser lines, to long period variables ($\dot{\mathrm M} \approx 10^{-6}$~M$_\odot$~yr$^{-1}$) where the main lines are more intense than that at 1612 MHz, and finally to OH/IR stars ($\dot{\mathrm M} \approx 10^{-4}$~M$_\odot$~yr$^{-1}$), where the 1612 MHz line is the brightest \citep{habing}. In OH/IR stars, the satellite line displays} doubly peaked profiles, with narrow components separated by $\sim$10--30 km s$^{-1}$. These profiles have been interpreted as arising from the approaching and receding sides of a spherically-symmetric, expanding envelope \citep{reid}. Emission of H$_2$O may  be doubly or singly peaked, with spectral components within the range marked by the two peaks of the OH emission at 1612 MHz. Since SiO masers form within the dust formation zone, they do not trace the AGB wind, therefore double peaks as those seen in expanding envelopes  are not expected in their line profiles. The appearance of SiO features at different velocities around the stellar value is instead due to the motion of the masing spots during a stellar variability cycle and indicates that these masers trace stellar pulsations, acting like a piston on the extended envelope.

OH and water maser spectra not following the regular double- (or single-) peak pattern, or H$_2$O maser emission extending outside the velocity range covered by the OH emission, are interpreted as evidence for non-spherical mass-loss processes \citep{gomez, deacon04}. The chronological sequence in which masers disappear as the star evolves to the PN phase is only valid for the case of spherically symmetrical mass loss: it will not stand if bipolar mass loss is present \citep{lewis}. This links the detection of H$_2$O and SiO masers after the tip of the AGB to non-spherical mass-loss processes.


A handful of astrophysical sources show hints for doubly-peaked SiO masers, as for example IRAS~$19312+1950$ and Orion Becklin-Neugebauer/Kleinmann-Low (BN/KL). IRAS~$19312+1950$ displays two peaks detected in the SiO v=2, J=$1-0$ line separated by $\sim$32 km~s$^{-1}$ and centred around the stellar velocity \citep{nakashima}. 
The source is probably a dust-enshrouded post-AGB star surrounded by an ambient cloud. 
In Orion BN-KL, the maser seems to be tracing a bipolar outflow from Source I,  thought to be a high-mass proto-stellar object \citep{reid07, goddi}.

 \subsection{Our target}
IRAS~$15452-5459$ (I15452) is a post-AGB star displaying an hourglass nebula, as seen with the Hubble Space Telescope (HST; \citealt{sahai}). All of the three maser species have been detected toward it, showing very peculiar features.\footnote{The target is indicated in the literature  also as d47.} Within the sample of stars observed by  \citet{deacon04}, this is the only source out of 88 with a four-peaked OH profile, and  one of only a few such sources known. 
 As reported by \citet{deacon07}, its water maser also shows several peaks spanning about 50 km s$^{-1}$, and SiO maser emission at 86 GHz was detected with Mopra, with two lines at approximately $-40$ and $-75$ km s$^{-1}$. The SiO data have a noise at best of 90~mK ($\sim$1~Jy/beam at Mopra), their velocities are uncertain to within $\pm1$~km~s$^{-1}$, and they lack absolute calibration \citep{deacon07}. 
The 86 GHz doubly-peaked SiO profile had previously been observed with the Swedish-ESO Submillimetre Telescope (SEST) in 1990 and with the Mopra in 1995, although the data were never published (te Lintel Hekkert 2010, private communication). 


In February 2010, we observed this source with the Atacama Pathfinder EXperiment (APEX) in the CO $J=3-2$ and $J=4-3$ lines at 345 and 461 GHz \citep{cerrigone}. We found wings of a fast molecular outflow with a velocity width of about 90 km~s$^{-1}$, larger than what observed in the OH maser in 2002/2003, thus either pointing to a fast velocity increase  or simply tracing a different CSE component. The source is likely to be rapidly evolving from the AGB into the post-AGB phase. 

A possible explanation for the SiO maser to be tracing out-flowing material may be that different SiO transitions originate in different regions. \citet{soria04, soria07} found that among the three SiO maser transitions routinely observed ($J=1-0$ $v=1$ and $v=2$ at 43 GHz, and $J=2-1$ $v=1$ at 86 GHz) the one at 86 GHz arises farther away from the central star than the others, concluding that   lines with low $J$ and high $v$ correspond to inner layers. 

To further investigate the CSE of this object and check for temporal variations, we observed I15452 in its OH, H$_2$O, and SiO masers at 1.6--1.7 GHz, 22.2 GHz, and 43 and 86 GHz, respectively. 

\section{Observations}
\label{sec:observations}
The observations were carried out with the Australia Telescope Compact Array in 2011 May and August.
In May, we observed the source in the 3~mm band (85--105 GHz), when the array was in its H214 configuration.  {This is a hybrid configuration, where two of the six antennas are located in a north spur instead of being aligned in east-west direction, for a better \textit{uv} coverage in a short time. The spacings between antennas ranged from 82 to 247 m.~\footnote{This does not take into account baselines with antenna 6, which is not equipped with a receiver at 3 mm.}}
Data in the bands at 16~cm (1.1--3.3 GHz), 15~mm (15--25 GHz), and 7~mm (29--51 GHz) were acquired in August, in 6B configuration. {This is an east-west linear configuration, with antenna spacings ranging between 214 and 5969 m.}

We carried out spectral-line observations of the OH maser transitions at 1612, 1665, 1667, and 1720 MHz, the water maser at 22.2 GHz, and the SiO maser and thermally excited lines near 43 and 86 GHz. The capability of the ATCA to simultaneously observe at two different frequencies within each observing band allowed us to check for emission in lines of HI (1.4 GHz), NH$_3$ (24 GHz), and the thermal line of SiO at 43 and 86 GHz (J=$1-0$ v=0 and J=$2-1$ v=0). The phase reference centre was at 15:49:11.380 in Right Ascension and $-$55:08:51.600 in Declination {(J2000)}.


To correct  atmospheric fluctuations, alternate scans on the target and on a nearby gain calibrator were observed. The complex-gain and bandpass calibrator was the AGN $1613-586$. 
The absolute calibration was obtained by bootstrapping the flux density of the complex-gain calibrator from that of $1934-638$ at all wavelengths except 3~mm, where the planet Neptune was used. The flux densities assumed for $1934-638$ were about 14.2 Jy at 16 cm, 0.75 Jy at 15~mm, and 0.28 Jy at 7~mm. \footnote{The specific values vary slightly depending on the central frequency of the single spectral window.} $1613-586$ was found to have flux densities of about 4.62 Jy (16 cm), 2.56 Jy (15 mm), 1.74 Jy (7 mm), and 2.12 Jy (3 mm). 
The uncertainties on the flux densities measured for the gain calibrator are dominated by the error on the absolute calibration, which can be regarded as 20\% at 3~mm and 10\% in the other bands.  For Neptune, which was slightly resolved by the array, the absolute scale was set by fitting to its observed visibilities a model of the planet with flux density at zero spacings of about 2.7 Jy.

Continuum data were also acquired with a bandwidth of 2 GHz in each observational band, with the constraint that the central frequency of each spectral window would fall within 1 GHz from the central continuum frequency. 

\subsection{Observations in the 3~mm band}
Our 3~mm observations were made on 2011 May 21, from 11:30:00 to 16:30:00 UT.  
One spectral window was placed on the SiO maser line at 86243.430 MHz ($J=2-1$, $v=1$) and another was centred on the frequency of the thermal SiO line at 86846.995 MHz ($J=2-1$, $v=0$). This allowed us to span 222 km~s$^{-1}$ with channel spacing of 0.11 km~s$^{-1}$ in each window.

Paddle measurements were carried out every 20 minutes, to correct for changes in the sky opacity, and pointing adjustments were calculated each hour on the complex-gain calibrator. The weather conditions during the run ranged from good to medium.

\subsection{Observations in the 15 and 7 mm bands}
The observations were carried out on 2011 Aug 27 from 03:00:00 to 11:00:00 UT, switching about every hour from one band to the other, to cover a range of hour angles as wide as possible in each band. The weather was very variable during the run and mostly mediocre. 
Because of the weather instability, the noise in the data {was slightly larger than expected}. Like at 3 mm, the pointing was adjusted once per hour by observing the complex-gain calibrator, while paddle measurements were not necessary, as the system temperature is continuously measured by default in bands other than 3 mm.

We centred one spectral window on the water maser, choosing a frequency of 22235.07985 MHz. 
The second  window in the same observing band was centred at the mean frequency of the two ammonia lines at 23694.506 and 23722.634 MHz. The channel width was 0.4 km~s$^{-1}$ and the width of each window was 809 km~s$^{-1}$.

At 7 mm, the two spectral windows were centred on the SiO maser line at 43122.080 MHz ($J=1-0$, $v=1$) and on the SiO thermal line at 43423.858 MHz ($J=1-0$, $v=0$). The channel spacing was 0.22 km~s$^{-1}$, spanning 442 km~s$^{-1}$.

\subsection{Observations in the 16 cm band}
Our centimetric run was carried out on 2011 Aug 26 from 03:00:00 to 07:00:00 UT and from 10:00:00 to 12:00:00 UT. To achieve high spectral resolution, a different set-up was used in this band, which provided us with up to 16 spectral windows for each of the two frequencies observable within the band. We centred a combination of 5  windows on the OH maser line at 1612.231, 5 on the line at 1665.402 MHz, and 6  on the line at 1720.530 MHz, thus spanning about 550 and 600 km~s$^{-1}$, respectively, with spacings of $\sim$0.1~km~s$^{-1}$. The 16 windows available at the second frequency were split between the OH line at 1667.359 MHz and the HI line at 1420.405752 MHz, spanning about 800 and 900 km~s$^{-1}$, respectively, with channels of $\sim$0.1~km~s$^{-1}$ width.

\section{Data reduction and results}
We reduced the data with the MIRIAD software package, which is the standard software for ATCA data, following the {recipe} in the MIRIAD User Guide. The raw data were loaded into MIRIAD with the task ATLOD performing automatic flagging of channels with known issues. The data were then corrected with the task ATFIX. This is mostly relevant for high frequency observations, to apply opacity and pointing corrections. This task also allowed us to load the improved antenna positions provided on the ATNF web site for our observations at 3 mm.

After flagging a few obvious outliers in the data points, we calibrated the bandpass on the secondary calibrator, then the complex-gain solutions were determined for $1613-586$ and the amplitude was scaled with values determined from our observations of the absolute  flux calibrator ($1934-638$ or Neptune),  selecting a time range at which the elevation of the secondary (phase) calibrator was as similar as possible to that of the absolute calibrator when the latter was observed. More flagging was then performed and gains re-calculated, until a satisfactory calibration of the data was achieved. The gains were then applied to the target and 3--4 iterations of phase-only self-calibration were performed, taking as a model a map obtained by averaging together the channels with the strongest signal.

Finally, data cubes were obtained by inverting the interferometer equation, cleaning the dirty cube, and then restoring the cleaned cube with a Gaussian beam. A summary of the beam sizes and channel spacing obtained is given in Table~\ref{beam}. In determining the weights to assign to each baseline when inverting the interferometer equation, a value of 0 was taken for the ROBUST parameter, thus balancing sensitivity and angular resolution. As the ATCA does not Doppler track, but records the instantaneous velocity of the observatory, the data were re-sampled in uniform velocity bins, when creating the cubes, which accounts for the relative motion between the target and the observatory.

\begin{table}\centering
\begin{tabular}{lcc}
\hline
\hline
Band  &  $\theta_\mathrm{B}$ &  Channel spacing \\
\hline
16 cm &  $8.0'' \times 4.8''$,  PA $31.2^\circ$ &  0.1 \kms \\
15 mm &  $1.3'' \times 0.26''$,  PA $11.7^\circ$ & 0.4 \kms\\
7 mm  &  $0.68'' \times 0.16''$,  PA $2.0^\circ$ & 0.2 \kms\\
3 mm &   $1.9'' \times 1.6''$,  PA $3.0^\circ$  & 0.1 \kms\\
\hline
\end{tabular}  
\caption{Beam and channel spacing obtained in the different observing bands.}
\label{beam}
\end{table}

\begin{figure*}
\centering
\includegraphics[width=0.95\textwidth]{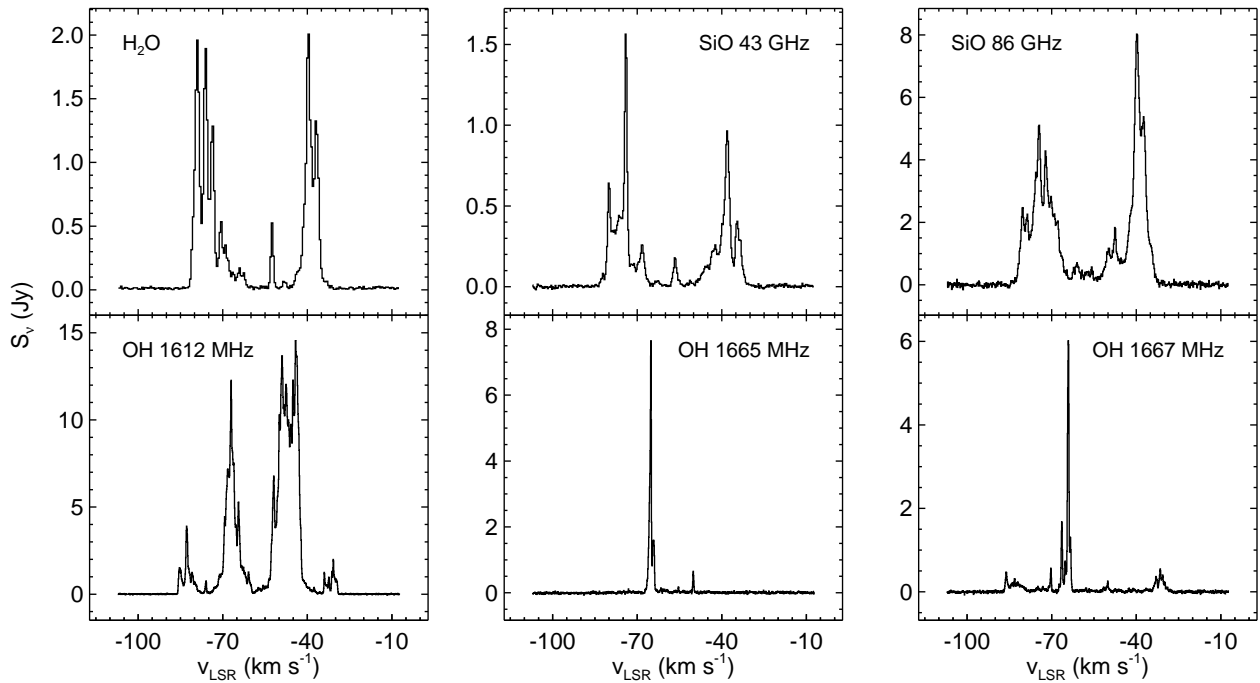}
\caption{Maser spectra of I15452 obtained in 2011 with the ATCA.}
\label{results}
\end{figure*}

\begin{figure}
\centering
\includegraphics[width=0.4\textwidth]{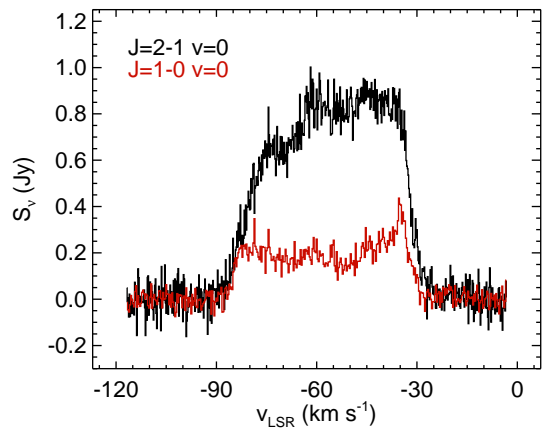}
\caption{Thermal lines of SiO at 86 GHz (black) and 43 GHz (red). At 43 GHz, the spectrum  is that observed on the shortest baseline.}
\label{thermalsio}
\end{figure}

After mapping the emission, we noted a slight offset of the target with respect to the pointing coordinates of about $-0.5''$ in right ascension and $0.3''$ in declination. We obtained the spectra of the targeted lines with the task UVSPEC as the vector average of the real part of the complex visibility at the off-set position, after re-sampling in velocity. These spectra are shown in Figures~\ref{results} and \ref{thermalsio}.

No emission was detected in the ammonia lines, with an rms noise of 11 mJy~beam$^{-1}$ over channels of 0.5 km~s$^{-1}$, nor in the HI line with an rms noise of 21 mJy~beam$^{-1}$ over channels of 0.1 km~s$^{-1}$. The continuum data did not lead to any detection over a 3$\sigma$ level, with rms noise of 0.42 mJy~beam$^{-1}$ at 2.01 GHz, 74 $\mu$Jy~beam$^{-1}$ at 23.7 GHz, 0.16 mJy~beam$^{-1}$ at 43.4 GHz, and 0.15 mJy~beam$^{-1}$ at 86.5 GHz. 

\begin{figure*}
\centering
\includegraphics[width=0.8\textwidth]{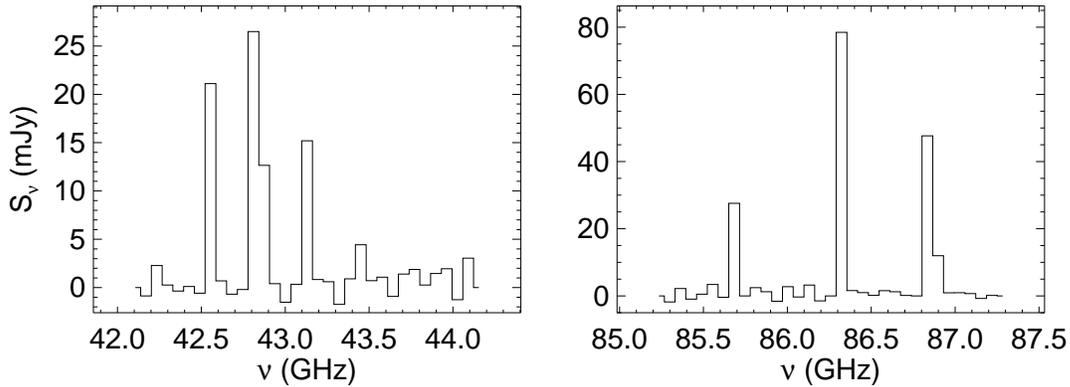}
\caption{Coarse (continuum) spectra observed in the 7 mm (\textit{left}) and 3 mm (\textit{right}) bands.}
\label{fig:cont}
\end{figure*}

In Figure~\ref{fig:cont}, we display the coarse (continuum data) spectra obtained at 7 and 3~mm, in which the thermal and maser SiO lines at 43.122,  43.424, 86.243, and  86.847 GHz can be seen. Other features are also detected. A third feature is observed between 85.6 and 85.7 GHz. Two more are seen at 7 mm: one between 42.5 and 42.6 GHz and another between 42.8 and 42.9 GHz. A higher resolution spectrum is needed for a proper identification of these lines or line overlaps. With the current data, we can just notice that the observed peaks occur at frequencies close to those of SiO lines with higher $v$. The rest frequencies for these  transitions would be about 42.519 GHz ($J=1-0$, $v=3$), 42.820 GHz ($J=1-0$, $v=2$), and 85.640 GHz ($J=2-1$, $v=2$).   

The velocity of the central star can be estimated from the OH maser line at 1665 MHz as $-57.7\pm0.1$ km s$^{-1}$.
{As clearly displayed by the profile of the line at 1612 MHz, four OH spectral peaks are detected. Taking as a reference the velocity of the central star, we can distinguish one pair of brighter peaks at velocities closer to the stellar value (roughly around $-47$ and $-67$ \kms), and a pair of weaker peaks detected at velocities further away from the stellar value (roughly $-33$ and $-83$ \kms). We will refer to the former as the \textit{slow} peaks and the latter as the \textit{fast} peaks.}

\begin{figure}
\centering
\includegraphics[width=0.45\textwidth]{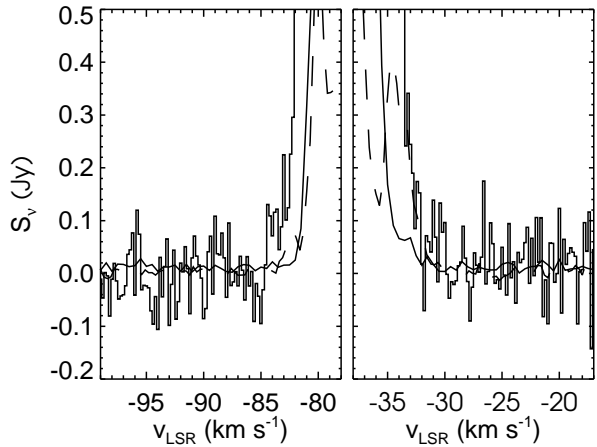}
\caption{Close-ups of the water and SiO maser lines at their most blue-shifted (\textit{left}) and red-shifted (\textit{right}) velocities. The SiO maser at 86 GHz is displayed as a histogram, that at 43 GHz as a dashed line, and the water maser as a solid line.}
\label{wings}
\end{figure}

From the OH line at 1665 MHz, we can estimate that the slow peaks trace material expanding at about 9~km~s$^{-1}$, while the expansion velocity derived from the fast peaks in the line at 1667 MHz is about 30~km~s$^{-1}$. The latter matches what we found in the satellite line, while the material generating the slow peaks in this feature has a broader range of velocities than in the main lines. As can be seen in Figure~\ref{wings}, the water and SiO masers span the same velocity range, from about $-85$ km~s$^{-1}$ to about $-30$ km~s$^{-1}$, corresponding to an expansion velocity of $\sim$27.5 km~s$^{-1}$. 

\section{The line profiles at different epochs}

\subsection{The OH masers}
\begin{figure*}
\centering
\subfigure[]{\includegraphics[height=0.22\textheight]{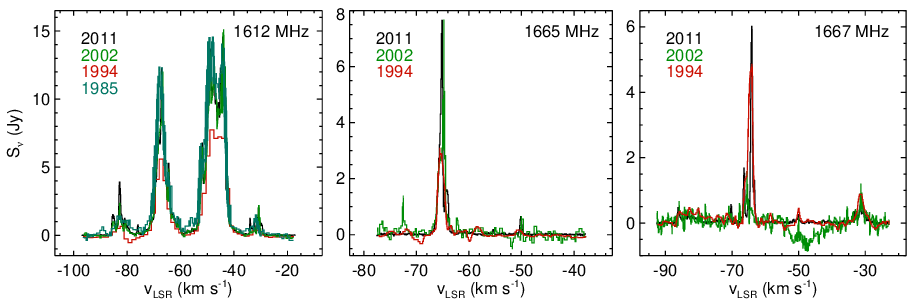}}
\subfigure[]{\includegraphics[height=0.22\textheight]{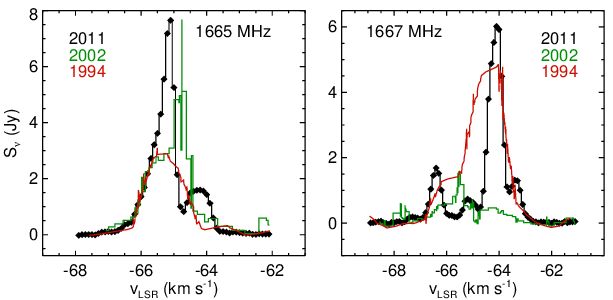}}
\caption{Comparison of OH maser lines observed at different epochs, with the full velocity range (\textit{a}) and close-ups of the brightest peaks of the two main lines (\textit{b}). In (\textit{b}) the data are displayed as a solid line (1994 data-set), a histogram line (2002 data-set) and a histogram with filled diamonds overlayed (our data-set).}
\label{oh_history}
\end{figure*}

The satellite line at 1612 MHz was observed in 1985 \citep{telintel}, 1994 \citep{sevenster}, and 2002 \citep{deacon04}, while the main lines were observed in 1994 \citep{caswell} and 2002 \citep{deacon04}. In Figure~\ref{oh_history}, we can see that the satellite line seems  remarkably stable over more than 15 years, with the exception of 1994, when it appears much weaker. However,  in 1994 the data were taken at a coarser velocity resolution ($\sim$1.5 km~s$^{-1}$) than those from other years, possibly leading to lower intensities due to spectral dilution. We note that for all epochs the maser emission components are spread over the same total LSR velocity range.  

A higher degree of variability is observed in the main lines. In 2002, the line at 1665 MHz displayed a secondary peak around $-73$ km~s$^{-1}$ that is not detected in our data. The feature at 1667 MHz was much brighter in 2011 {than in 2002} and clearly shows the fast secondary peaks seen in the satellite line, which were too weak to be clearly detected in 2002. 

\begin{table}\centering
\begin{tabular}{lclclc}
\hline
\hline
\multicolumn{2}{c|}{OH 1612 MHz} &\multicolumn{2}{c}{H$_2$O}&\multicolumn{2}{c}{SiO 86 GHz}\\
Year & $\int \mathrm{S_\nu} dv$ & Year & $\int \mathrm{S_\nu} dv$& Year & $\int \frac{\mathrm{S_\nu}}{S_{max}} dv$\\
     &  Jy \kms &     &  Jy \kms &  & \kms \\
\hline
1985 &        $183\pm18$ & 2003  &  $36\pm4$  & 1990 & $11\pm2$ \\
1994 &        $91\pm9$   & 2003  &  $35\pm3$  & 1995 & $6\pm1$ \\
2002 &        $148\pm15$ & 2004  &  $11\pm1$  & 1995 & $11\pm2$ \\
2011 &        $156\pm16$ & 2011  & $21\pm2$   & 2011 & $11\pm2$ \\
\hline
\end{tabular}
\caption{Integrated flux densities of the maser lines at different epochs.}\label{integrated}
\end{table}

\subsection{The water maser}
\begin{figure}
\centering
\includegraphics[width=0.4\textwidth]{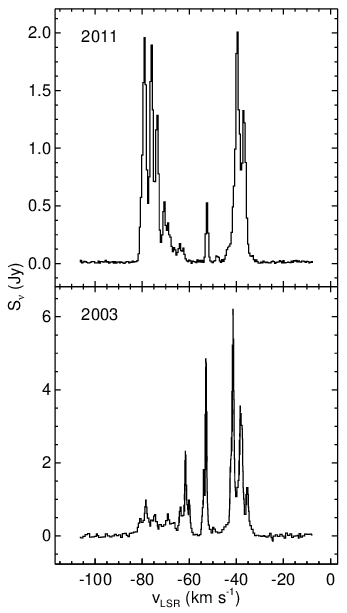}
\caption{Water maser in I15452 at two different epochs: our data-set from 2011 and a 2003 data-set digitised from \citet{deacon07}.}
\label{water}
\end{figure}

Water masers are known to be variable on time scales of some months. \citet{deacon07} observed the water masers in I15452  in August and November 2003 and in May 2004. They noticed that while the spectral features were stable in velocity, their relative intensities displayed major time variations. We digitised their spectra and display in Figure~\ref{water}  their data
from August 2003 and ours. 
Our spectrum from 2011 displays the same velocity span as that from  2003, but the overall spectral profile has changed and it has substantially dimmed, its integrated flux being about half of what found by \citet{deacon07}. In our data, we clearly see a double-peak profile that was not evident in the old data-set, where the red-shifted peak {around $-37$ \kms} seen in 2011 is present, while the blue-shifted one {around $-77$ \kms} is weak. What is instead brighter in the data from 2003 is a  doubly-peaked feature centred around the stellar velocity and about 15 km~s$^{-1}$ wide, {with peaks roughly around $-53$ and $-63$ \kms}. In 2011, only the red-shifted peak of this spectral component is detected. 

The overall variation in line flux of our 2011 data seems to range within the previously observed one (Table~\ref{integrated}).

Observations at higher angular resolution are necessary to interpret the data, but it seems reasonable to see the central, slow component as the water analogue of the central peaks seen in OH,  while the fast peaks  trace a different structure.

\subsection{The SiO maser}
\begin{figure}
\centering
\includegraphics[width=0.45\textwidth]{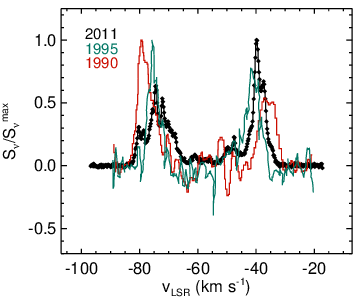}
\caption{Overlay of the SiO maser line at 86 GHz  obtained with the SEST in 1990 (red histogram line), with Mopra in 1995 (aquamarine solid line), and our data (black with filled diamonds).}
\label{old+new}
\end{figure}

Observations of the SiO maser at 86 GHz were taken at different epochs and with different telescopes. In Figure~\ref{old+new}, we have plotted our maser data at 3~mm over previous Mopra and SEST spectra, which were recovered directly from an on-line image\footnote{The image can be found at \url{http://www.atnf.csiro.au/people/plintel/groups/lsse/members/lintel/15452.html}}. The old data were acquired in 1990 with the SEST and in August and November 1995 with Mopra. The spectra lack absolute calibration and  were smoothed after digitalisation, to improve their signal-to-noise ratio. 

SiO masers are known to be highly variable in intensity, though spanning a stable range of velocities. An inspection of Figure~\ref{old+new} seems to indicate that from 1990 to 1995 the width at zero intensity had decreased, while there may be an increase from 1995 to 2011. Anyway, it is difficult to draw any conclusions, because of the limited quality of the Mopra and SEST data-sets. After peak normalisation, the integrated line flux density is quite constant with time, within a 20\% absolute uncertainty (Table~\ref{integrated}). 

\section{The spatial distribution of the maser emission}
To identify the maser features and estimate their peak positions on the sky, we re-sampled the spectra of the OH satellite line and of the SiO maser at 7 mm to a velocity resolution of 0.5 \kms and then fitted the emission peak in each channel with a 2D Gaussian, if the emission was detected in three consecutive channels with at least S/N$=6$. The water maser was not re-sampled to a coarser velocity resolution. 
Because of their particularly elongated beam, the  data at 7 and 15 mm were not only re-sampled, but also convolved with a round Gaussian beam, whose radius was the geometric mean value of the major and minor semi-axes of the original beam. The sky position of each maser feature was then calculated on the  mean of each set of consecutive channels. The positional uncertainty of each maser spot depends on the beam size and the S/N of the map and  is given by $0.5\,\theta/\,(\mathrm{S}/\mathrm{N})$. {Brighter spectral components therefore have smaller positional uncertainty than weak ones. The uncertainties range from 5 to 15 mas at 7 mm, from 2 to 25 mas at 15 mm, and from 3 to 230 mas at 16 cm, with average values of $7\pm2$ mas, $7\pm5$ mas, and $40\pm43$ mas, respectively. }

\begin{figure}
\centering
\includegraphics[width=0.4\textwidth]{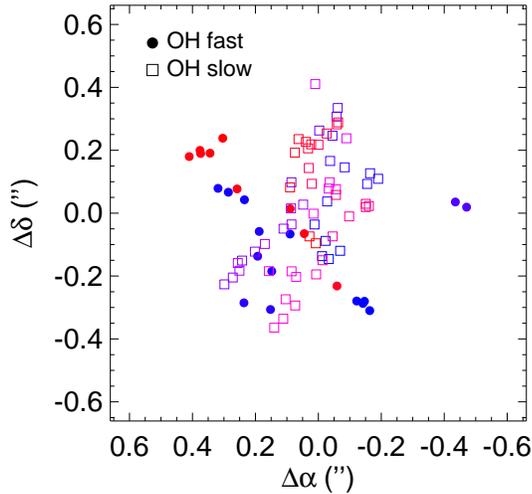}
\caption{Spatial distribution of the OH maser spots (1612 MHz). Open squares indicate spots corresponding to velocities between $-75$ and $-35$ km s$^{-1}$ (slow peaks), filled circles indicate velocities out of this interval and ranging between $-85$ and $-29$ km s$^{-1}$  (fast peaks). The velocity is colour coded ranging from $-85$ \kms (blue) to $-29$ \kms (red).}
\label{oh}
\end{figure}

In Figure~\ref{oh}, we show the results of this procedure for the OH line at 1612 MHz, the water maser, and the SiO maser at 43 GHz. Although our beam sizes are large compared to the sizes of the emitting regions, some general conclusions can be drawn.  Different spatial distributions are found for the features corresponding to the slow and  fast peaks of the OH maser and  
the two directions along which the spots are seen  are roughly perpendicular. 
This is {a significant advance in understanding the peculiar} OH spectral profile. The fast and slow components seem to trace different structures of the CSE. 

\begin{figure*}
\centering
\includegraphics[width=0.4\textwidth]{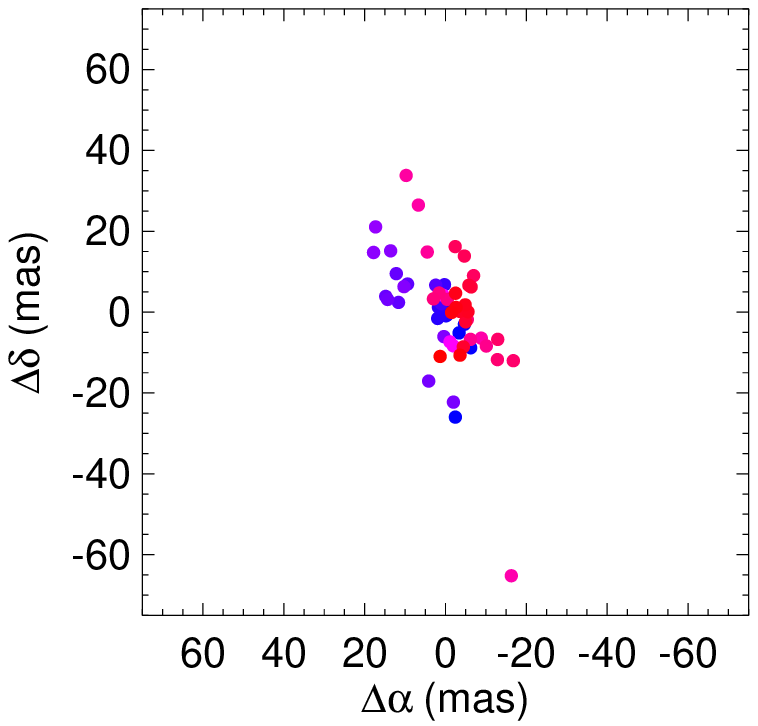}\hspace{1.5cm}
\includegraphics[width=0.4\textwidth]{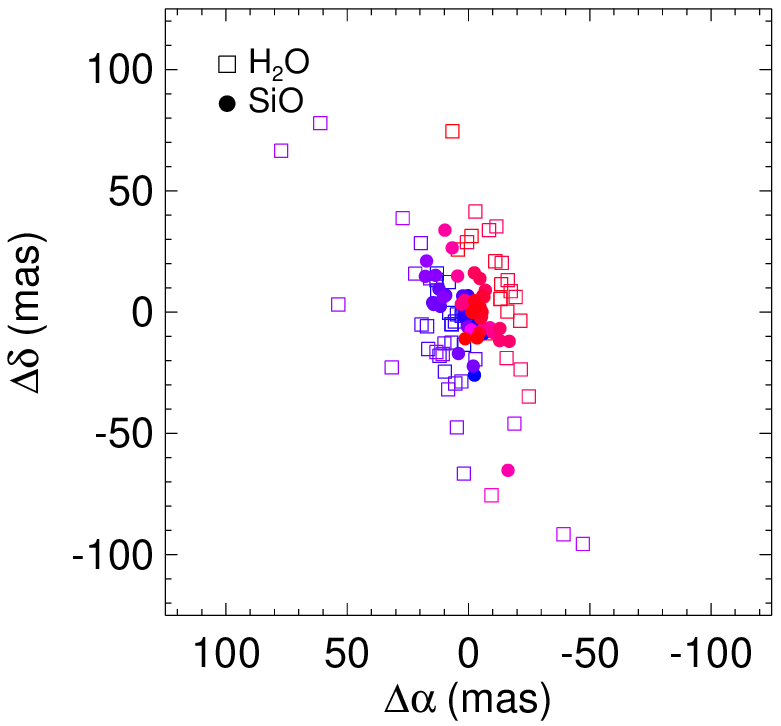}
\caption{Spatial distribution of the emission peaks of  7~mm SiO masers (\textit{left}) and the same peaks overlayed on the water maser peaks (\textit{right}) relatively to an arbitrary common centre. The velocities range from $-82$ (colour-coded as blue) to $-32$ km s$^{-1}$ (colour-coded as red)  and the average position uncertainties are $\sim$6 mas for SiO and $\sim$5 mas for water masers.}
\label{water+sio}
\end{figure*}

As can be seen in Figure~\ref{water+sio}, the SiO masers display an hourglass shape, while the water spots trace a larger region surrounding that of SiO. 
Higher angular resolution is necessary to allow for a thorough investigation of these distributions. What we detect is  the inner region of the outflow that at larger distances from the central star is creating the nebular shape observed with the HST.

Both the water and SiO maser spots display a velocity gradient  in the direction perpendicular to the main axis of the hourglass, along which no clear gradient is seen. The SiO spots are distributed over an area of about $68 \;\mathrm{mas}\times 25\; \mathrm{mas}$. At the assumed distance to I15452 of 2.5 kpc, 
  T$_\star=3500$ K, and L=8500 L$_\odot$, this would mean $53 \; \mathrm{R}_\star \times 144 \;\mathrm{R}_\star$. 

\begin{figure}
\centering
\includegraphics[width=0.45\textwidth]{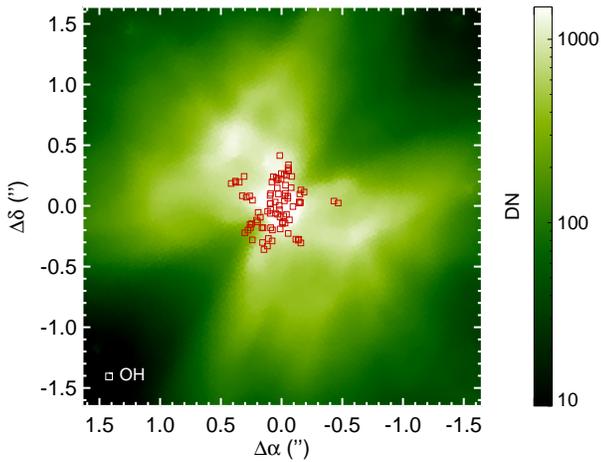}
\caption{The central region of I15452 as seen with NICMOS@HST. Open squares indicate OH (1612 MHz) maser spots.} 
\label{hubble}
\end{figure}

We compare in Figure~\ref{hubble} the OH maser spots to the emission observed with the HST at near-IR wavelengths shown by \citet{sahai}. The data have been {centred around an arbitrary common reference position. 
This comparison suggests that the slow OH spots may actually be spatially linked to the inner regions of the two lobes seen in the HST images.} 

\begin{figure}
\centering
\includegraphics[width=0.25\textwidth]{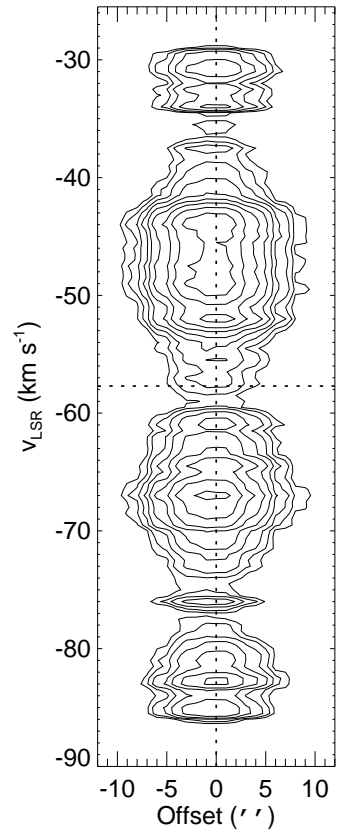}
\caption{Position-velocity diagram of the OH line at 1612 MHz along the direction where the slow spots are observed. The central position and the assumed stellar velocity are displayed as dotted lines. }
\label{pvplot}
\end{figure}

{We constructed a position-velocity diagram (Figure~\ref{pvplot}) for the 1612 MHz OH line.} The offsets are calculated along a direction approximately aligned with the area where the slow peaks from Figure~\ref{oh} are seen. We also checked the result for cuts along different directions and no major changes were observed.  {Although position-velocity diagrams {can be} powerful tools to analyse the regions where lines arise, the lack of angular resolution hampers this type of analysis in our case. Besides} some hints for slight asymmetries, for example around $-67$~\kms, the plot does not show any trend of the velocity as a function of the position. 

\section{Thermal lines of SiO}
We detected for the first time in this source thermal lines of SiO at 43 and 86 GHz. The spectra are displayed in Figure~\ref{thermalsio}. The profiles of the thermal lines at 3 and 7 mm are different. At 3 mm, we see a flat-topped line, while at 7 mm  {a double-horned shape is} observed. Both profiles are typical of optically-thin lines, but the former implies that the source is not angularly resolved, while the opposite is true for the latter.  At 7~mm, the ATCA provided enough angular resolution to partly resolve the region emitting the thermal line {(deconvolved size of $1.8''\times 0.3''$ at PA$=16.8^\circ$)},  shown in Figure~\ref{thermalsiomap}. Unfortunately, as snapshot observations were performed, the beam resulted to be strongly elongated in a direction close to the nebular axis.

In AGB stars, SiO maser conditions are met only within a few R$_\star$ from the stellar surface, where there may be turbulence, but  no outgoing wind. Conversely, thermal emission from SiO has been observed farther away from the central star both in O-rich and C-rich sources, with expansion velocities 20\% smaller than  {those} derived from CO lines \citep{schoier, delgado, ramstedt}. This difference in velocity is interpreted as due to SiO depletion into dust grains.  {This makes SiO in AGB stars a tracer of the inner layers, where there is no wind or this is still accelerating.}

\begin{figure*}
\centering
\includegraphics[width=0.9\textwidth]{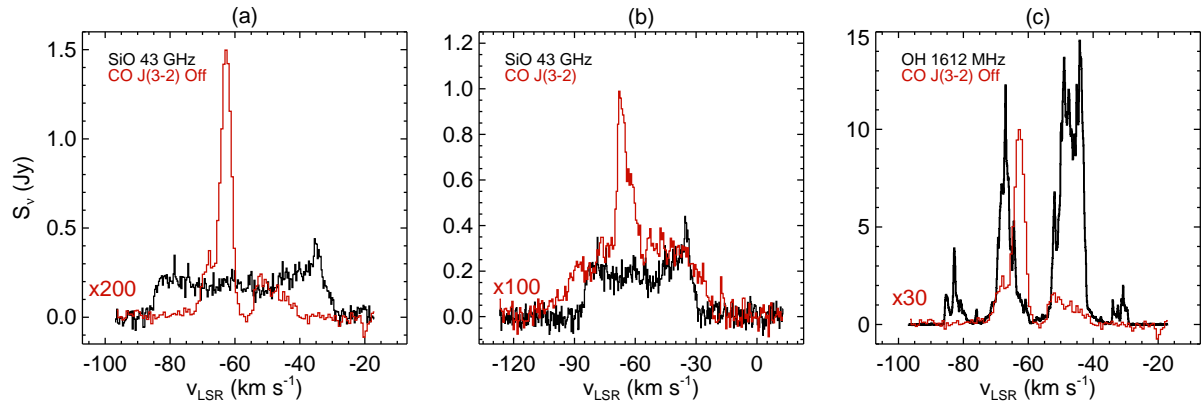}
\caption{Overlay of CO, SiO, and OH lines. The CO spectra are to be multiplied by the indicated factors.}
\label{sio_co}
\end{figure*}

\begin{figure}
\centering
\includegraphics[width=0.4\textwidth]{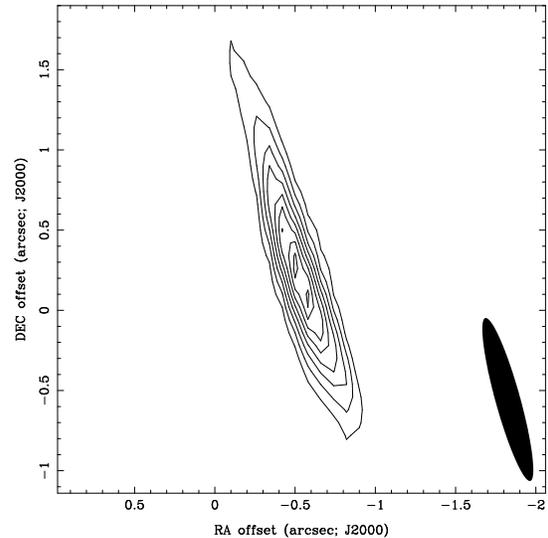}
\caption{Integrated intensity map of the thermal line of SiO at 43 GHz, after binning each two pixels in both right ascension and declination. The contours are at ($-3$, 3, 4, 5, 6, 7, 8, 9, 10)$\sigma$, where $\sigma= 0.4$ Jy km s$^{-1}$ and the reference position is at our pointing coordinates ($\S$ \ref{sec:observations}). The beam is displayed in the bottom right corner.}
\label{thermalsiomap}
\end{figure}

We can compare the velocity range derived from our SiO lines to that from CO reported by \citet{cerrigone}.  {The comparison is illustrated in Figure~\ref{sio_co}, where we display the SiO thermal line at 43 GHz, the CO J=$3-2$ line, and the OH satellite maser line. The \lq\lq off\rq\rq~spectrum refers to a pointing off-set from the central coordinates by 10$''$ \citep[for details]{cerrigone}.} The CO line shows a complex profile with a double peak overlayed on a fast broad feature  {(plot \textit{b} of Figure~\ref{sio_co})}. The central two peaks of CO    {(plot \textit{a} of Figure~\ref{sio_co})} span the same velocity range as the central peaks of the OH maser,  {as shown in plot \textit{c} of Figure~\ref{sio_co}} ($\Delta v= 28$ km s$^{-1}$). To set our SiO thermal lines in the same picture as in the AGB stars where these have been detected, we can  take as CSE expansion velocity that derived from the slow features in the profiles of the CO and OH lines. Unlike in AGB stars, we find that the expansion velocity from the SiO lines is about 1.8 times that of the expanding CSE.  {This implies that the SiO in our target is not tracing the inner wind acceleration region.}

 {SiO is considered as a reliable tracer of shocked gas in many astrophysical environments and its enhancement in the gas phase has  been observed for example in the bipolar outflows of young stellar objects. In fact, the propagation of a shock can remove Si from the grains by sputtering \citep{martin} or evaporation from the grain mantles \citep{gusdorf}. Silicon is thus released into the gas phase. In our target, the detection of thermal and maser lines of SiO with velocities larger than that of the CSE can then be explained by the pro\-pa\-gation of a shock, possibly as a consequence of a jet acting on the CSE. Since the shocked gas can move faster than the wind, this would explain why our SiO lines trace larger velocities.}


\section{Discussion}
I15452 belongs to a group of OH masing stars classified by \citet{sevenster} as LI, because they lie on the left of the evolutionary sequence on the IRAS colour-colour diagram \citep{vanderveen}, while \lq\lq standard\rq\rq~post-AGB stars lie on the right side (RI sources).   LI stars have higher outflow velocities and mass-loss rates and they may loop back to the left of the evolutionary sequence because of an early AGB termination, which would also cause them to host OH masers for a much longer time than RI stars \citep{sevenster}.

I15452 displays all the typical masers observed in AGB stars, yet its hourglass nebula and the velocities derived from the line profiles indicate that it has started its post-AGB evolution. 
A peculiarity of this star is the profile of its OH maser line at 1612 MHz with four peaks. {For another source known to display a similar profile, it was proposed that this may be due to a double-cone expanding envelope \citep{chapman}. 
Other hypotheses include an expanding and rotating disc \citep{grinin} and an elliptical envelope observed at a certain viewing angle \citep{bowers, collison}.}

Our high-angular resolution observations indicate that the four peaks are also present in the main line at 1667 MHz and that the slow peaks are roughly aligned along a direction  perpendicular to that of the fast ones. 

The spatial distribution of the OH masers that we find is compatible with the fast and slow peaks being related to different structures. 
The fast peaks trace the fast outflow/jet from the central star causing the broad CO features observed by \citet{cerrigone}. The CO-derived outflow velocity (v$_{exp}= 45\;$\kms) can be interpreted as the terminal velocity of this outflow, while the velocity derived from the fast OH peaks (v$_{exp}=30\;$\kms) would simply reflect the different excitation conditions of the maser line, which is tracing higher densities than CO, therefore  arises in regions where the outflow has not yet reached its terminal velocity. {The comparison of the OH maser spots with the HST near-IR images suggests that the slow peaks would instead arise from the inner walls of the two nebular lobes.}

In water-fountain nebulae, water masers arise from the tips of a collimated outflow, where this impinges on the remnant AGB wind. In I15452, the water masers are found to surround the region where the SiO masers are observed. Like the overall nebula, the SiO masers display an hourglass distribution, but at much smaller spatial scale. They are also characterised by a velocity gradient orthogonal to the main axis of the hourglass, indicative of a rotation around the main axis. The hourglass shape may be explained if we consider that the emission arises at the interaction  {region} between the fast wind and the torus observed by \citet{sahai}, or a smaller-scale structure, like a disc. An explanation for the orthogonal velocity gradient would be that the gas is tracing the rotation of the circumstellar structure, torus or disc, with still little acceleration along the nebular axis.

A similar velocity gradient has been observed, for example, in a strongly bipolar nebula as the PN M 2-9. In this object, knots  rotate around the main nebular axis and the rotation has been interpreted in terms of a precessing jet, with the knots being the interaction regions between the jet and the ambient gas \citep{doyle}. A precessing jet has also been invoked in IRAS $16342-3814$, to explain the corkscrew pattern seen in its bipolar envelope. A se\-cond possibility is therefore that a similar process at smaller scales may be causing the velocity gradient observed in the SiO and water masers in I15452, where the single condensations/knots would not be resolved by our observations. 

The detection of the thermal line of SiO indicates a strong enhancement of Si in the gas phase. The emission is partly resolved by our snapshot observations and appears roughly in the same direction of the nebular axis. 
 As such enhancements of Si are explained by sputtering of dust grains, this confirms that the detection of the SiO masers cannot be related to the typical AGB scenario, but is due to asymmetric mass loss and shock propagation. 

\section{Summary and conclusions}
{  We have presented spectral lines of the three chemical species typically found to display masers in evolved stars. Previous maser observations toward I15452 were available and different degrees of variability are found for different lines. While the main lines of OH display strong variability from an epoch to another, the maser at 1612 MHz is quite stable over a time of several years. The water maser is also found to show variability.

Our observations confirm the previous detection of doubly-peaked SiO maser profiles at 3~mm and a similar shape is found at 7~mm. We also detected thermally excited lines of SiO around both 43 and 86 GHz, for the first time in this source.

The analysis of the positions of the maser emission centroids leads us to conclude that the two pairs of peaks displayed by the satellite OH line arise from different structures rather than being due to projection effects from one emitting object. A similar analysis finds that the SiO centroids follow an hourglass spatial distribution.

Finally, the comparison of the velocity ranges spanned by different features, including the CO lines presented in \citet{cerrigone}, suggests that the outflow causing the CO emission may also be  traced, at higher densities, by the fast OH and ultimately by the SiO, whose presence is explained by shock propagation.}

\section*{Acknowledgements}
The Australia Telescope Compact Array is part of the Australia Telescope National Facility which is funded by the Commonwealth of Australia for operation as a National Facility managed by CSIRO. L.C. acknowledges support from the Spanish Consejo Superior de Investigaciones Cient{\'{\i}}ficas  through the JAE Doc programme.

\bsp

\label{lastpage}

\end{document}